%% file: main.tex
\documentclass[10pt,conference]{IEEEtran}
\IEEEoverridecommandlockouts
\usepackage{cite}
\usepackage{amsmath,amssymb,amsfonts}
\usepackage[acronyms]{glossaries}
\usepackage{algorithmic}
\usepackage{graphicx}
\usepackage{subcaption}
\usepackage{textcomp}
\usepackage{xcolor}
\usepackage{footnote}
\usepackage{soul}
\usepackage{url}
\newcommand {\Define} {\stackrel {\Delta} {=}  }
\makeatletter
\newcommand{\ostar}{\mathbin{\mathpalette\make@circled *}}
\newcommand{\make@circled}[2]{%
  \ooalign{$\m@th#1\smallbigcirc{#1}$\cr\hidewidth$\m@th#1#2$\hidewidth\cr}%
}
\newcommand{\smallbigcirc}[1]{%
  \vcenter{\hbox{\scalebox{0.77778}{$\m@th#1\bigcirc$}}}%
}
\makeatother

\def\BibTeX{{\rm B\kern-.05em{\sc i\kern-.025em b}\kern-.08em
    T\kern-.1667em\lower.7ex\hbox{E}\kern-.125emX}}

\input{acronyms}

\begin{document}


\title{Over-the-Air Transmission of Zak-\gls{otfs} on mmWave Communications Testbed \vspace{-5mm}
\thanks{}}



 \author{
   \IEEEauthorblockN{Sabarinath Ramachandran\IEEEauthorrefmark{2}\textsuperscript{\textsection}, Venkatesh Khammammetti\IEEEauthorrefmark{1}\textsuperscript{\textsection}, Prasanthi Maddala\IEEEauthorrefmark{2}, \\ Narayan Mandayam\IEEEauthorrefmark{2}, Ivan Seskar\IEEEauthorrefmark{2},  and Robert Calderbank\IEEEauthorrefmark{1}}
\IEEEauthorblockA{\IEEEauthorrefmark{1}\textit{Electrical and Computer Engineering Department, Duke University}, Durham, NC, USA \\ \IEEEauthorrefmark{2}\textit{WINLAB, Rutgers University}, North Brunswick, NJ, USA  \\ sr1520@scarletmail.rutgers.edu, venkatesh.khammammetti@duke.edu, 
prasanti@winlab.rutgers.edu, \\
narayan@winlab.rutgers.edu,  seskar@winlab.rutgers.edu, and robert.calderbank@duke.edu\vspace{-1.2em}
   }
}

\maketitle

\begingroup\renewcommand\thefootnote{\textsection}
\footnotetext{These two authors contributed equally to this work.}
\endgroup

\begin{abstract}
Millimeter-wave (mmWave) communication offers vast bandwidth for next-generation wireless systems but faces severe path loss, Doppler effects, and hardware impairments. Orthogonal Time Frequency Space (OTFS) modulation has emerged as a robust waveform for high-mobility and doubly dispersive channels, outperforming OFDM under strong Doppler. However, the most studied multicarrier OTFS (MC-OTFS) is not easily predictable because the input-output (I/O) relation is not given by (twisted) convolution. Recently, the Zak-transform-based OTFS (Zak-OTFS or OTFS 2.0) was proposed, which provides a single-domain delay–Doppler (DD) processing framework with predictable I/O behavior. This paper presents one of the first over-the-air (OTA) demonstrations of Zak-OTFS at mmWave frequencies. We design a complete Zak-OTFS-based mmWave OTA system featuring root-raised-cosine (RRC) filtering for enhanced DD-domain predictability, higher-order modulations up to 16-QAM, and a low-overhead preamble for synchronization. A comprehensive signal model incorporating carrier frequency offset (CFO) and timing impairments is developed, showing these effects can be jointly captured within the effective DD-domain channel. Experimental validation on the COSMOS testbed confirms the feasibility and robustness of Zak-OTFS under realistic mmWave conditions, highlighting its potential for efficient implementations in beyond-5G and 6G systems. 
 \end{abstract}
\glsresetall


\section{Introduction}
Growing demand for higher data rates, ultra-low latency, and pervasive connectivity is pushing wireless systems toward ever-higher carrier frequencies and more challenging propagation regimes. \gls{mmwave} bands, with their abundant spectrum, are widely recognized as a key enabler for next-generation wireless systems (5G, beyond-5G, 6G) \cite{mmwave5G_2}. However, mmWave links face significant hurdles: severe path loss, susceptibility to blockages, hardware impairments (e.g., phase noise), and the pronounced effects of mobility (Doppler) and multipath (delay spread) \cite{mmwave_challenges_1,mmwave_challenges_2}. In the presence of high mobility, conventional modulation schemes such as OFDM struggle due to  \gls{ici} and time selectivity of the wireless channels\cite{OFDM_ICI_3}. 

\gls{otfs} ((also termed as \gls{mcotfs})) was introduced as a promising alternative for doubly dispersive channels. In \gls{otfs}, information symbols are mapped onto the \gls{dd} domain, rather than in the time or frequency domain \cite{Hadani2017}. \gls{mcotfs} is an OFDM overlay where the signals are formed in the \gls{dd} domain, then mapped to \gls{tf}. The essential benefit is that under many channel conditions, the rapidly time-varying, frequency-dispersive channel in \gls{tf} becomes a nearly time-invariant effective channel in \gls{dd}, enabling more robust equalization and improved performance in high Doppler scenarios \cite{Hadani2015OTFSWhitePaper}. 

Over the last several years, a growing number of works have explored the application of \gls{otfs} specifically in mmWave settings, both from theoretical and implementation perspectives. 
\gls{otfs} is a candidate waveform for mmWave systems due to its inherent robustness to Doppler and phase noise encountered at higher carrier frequencies \cite{Hadani2015OTFSWhitePaper,Samsung_mmwave,OTFS_mmwave_2025}. The comparative simulations in \cite{Hadani2015OTFSWhitePaper} indicate that \gls{otfs} can outperform \gls{ofdm} in many mobility regimes, especially where Doppler-induced \gls{ici} becomes significant. Later, researchers in \cite{Samsung_mmwave} evaluated \gls{otfs} over a realistic 5G NR \gls{mmwave} transmit-receive chain using standardized channel models (TDL, CDL), assessing BER and complexity tradeoffs versus \gls{ofdm}. Their results showed that, under certain conditions (e.g. high speed train, 500 km/h), \gls{otfs} can offer $~10$ dB SNR gain over \gls{ofdm} for 64-QAM in systems with large delay spread. In a more recent work, the authors in \cite{OTFS_mmwave_2025} compared Single Carrier, \gls{ofdm}, and \gls{otfs} in \gls{mmwave} Multi-Connectivity Downlink Transmissions, explored multiple access point downlink scenarios with imperfect synchronization, and showed that \gls{otfs} significantly outperforms single-carrier and \gls{ofdm} in pragmatic capacity metrics, albeit with additional complexity. Also, the authors in \cite{OTFS_mmwave_2023_1} proposed a low-complexity hybrid precoding scheme for OTFS-based mmWave air-to-ground systems by formulating the \gls{dd} domain channel and designing corresponding hybrid digital–analog precoders. Similarly, \cite{OTFS_mmwave_2023_2} introduced Delay–Doppler Alignment Modulation (DDAM-OTFS), which applies per-path delay and Doppler compensation using large antenna arrays and channel sparsity in mmWave/THz bands. These studies demonstrate that OTFS, combined with beamforming and precoding strategies, is a promising waveform for mmWave communications. In recognition of real RF impairments at mmWave, the authors in \cite{Surabhi2019OTFSPhaseNoise} integrate oscillator phase noise into the delay–Doppler domain channel model (for a 28 GHz system) and employ a message-passing detector. Their results indicate that OTFS maintains robustness under phase noise, and can outperform OFDM by orders of magnitude in BER under comparable conditions.

While the \gls{mcotfs} architecture is widely studied, \gls{otfs}  formulations based on the \gls{zotfs} have gained traction owing to implementation benefits and \gls{io} predictability compared to \gls{mcotfs} \cite{Saif2_base,Saif2_chan}. \gls{mcotfs} increases both computational burden and implementation latency, especially when mapping between \gls{tf} and \gls{dd} domains. In contrast, \gls{zotfs} process as a single-domain operation in the \gls{dd} domain, producing a direct and predictable \gls{io} relation, which improves implementation efficiency, lower hardware complexity, and enhanced \glspl{io} determinism under practical conditions.

However, while \gls{zotfs} has shown strong theoretical potential, its practical implementation in mmWave systems has not been explored. Existing studies \cite{Zakotfs_sub6ghz, Zakotfs_thz} have examined \gls{zotfs} realizations at sub-6 GHz and THz frequencies, yet no comprehensive work has addressed \gls{mmwave} operation. Moreover, these investigations overlook critical real-world hardware impairments, including \gls{cfo} and timing misalignment. 

In this paper, we employ \gls{zotfs} (OTFS 2.0) \cite{Saif2_base} as the physical-layer waveform and perform \gls{ota} implementaion over \gls{mmwave} frequencies. The main contributions of this work are as follows:
\begin{itemize}
    \item \textit{Experimental Validation}: Developed and demonstrated a \gls{mmwave} \gls{ota} system employing \gls{zotfs} as the physical-layer waveform, with \gls{ota} experiments conducted at pedestrian walking speeds, representing one of the first experimental validations of \gls{zotfs} at \gls{mmwave} frequencies on the COSMOS testbed \cite{cosmos2025}.
    \item \textit{System Design Improvements}: Designed an enhanced Zak-OTFS–based OTA architecture that improves input–output (I/O) predictability through the use of \gls{rrc} filters instead of traditional sinc filters. The proposed setup also supports higher-order constellations (up to 16-QAM) and employs a low-overhead preamble for signal synchronization, outperforming prior implementations \cite{Zakotfs_sub6ghz, Zakotfs_thz}.
    \item \textit{Signal Modeling with Hardware Impairments}: Formulated the Zak-OTFS received signal model incorporating hardware impairments such as carrier-frequency offset (CFO) and timing offsets, and demonstrated that these effects can be jointly estimated within the effective channel response.
\end{itemize}

In the remainder of the paper, we present the \gls{zotfs} system model in Section \ref{section:Transceiver Design}, Section \ref{section:Otfs implementation} then describes our contribution to received signal formulation, improved \gls{ota} implementation of \gls{zotfs} and the \gls{mmwave} experimental setup. Section \ref{section:results} presents the performance evaluations obtained with our experimental setup over \gls{mmwave} frequencies.

\section{Zak-\gls{otfs} System Model}
\label{section:Transceiver Design}
Consider a \gls{dd} frame of fundamental periods $\tau_p$ and $\nu_p$ along the delay and Doppler domains, respectively, such that $\tau_p \nu_p = 1$. The delay and Doppler periods are sub-divided into $M$ and $N$ eqaul parts such that each \gls{dd} resolution is $\frac{\tau_p}{M}$ along the delay domain and $\frac{\nu_p}{N}$ along the Doppler domain. 
\subsection{Transmitter Signal Processing}
\label{subsection:zak_tx}
\begin{figure}
    \centering
    \includegraphics[width=\linewidth]{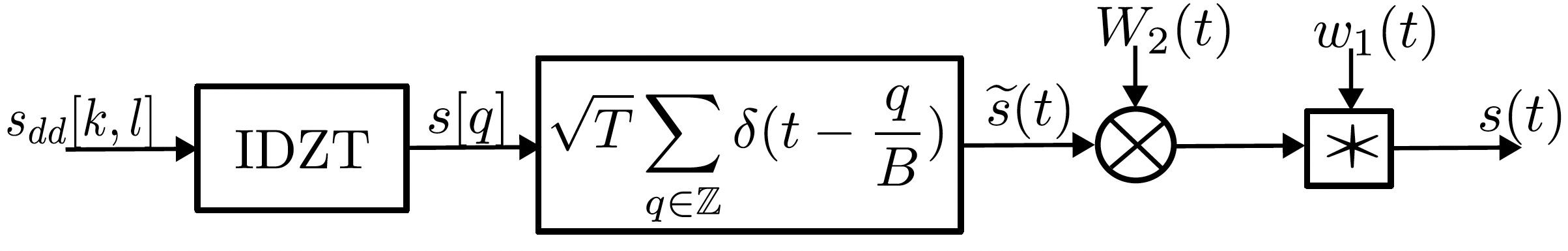}
    \caption{Zak-OTFS transmitter}
    \label{fig:zak_tx}
    \vspace{-6mm}
\end{figure}
The transmitter signal processing is shown in Fig. \ref{fig:zak_tx}. Let $s[k, l], k=0,1, \cdots, M-1, l=0,1, \cdots, N-1$ denote the $MN$ information symbols multiplexed in a \gls{dd} frame. At the transmitter, $s[k, l]$ are embedded into a quasi-periodic discrete \gls{dd} domain signal\footnote{Quasi-periodic extension is required to apply Zak transform (refer \cite{Saif2_base} for more details)} with periods $M$ and $N$ along the discrete delay and Doppler axis, i.e.,
\begin{equation}
s_{d d}[k+n M, l+m N]=e^{j 2 \pi n \frac{l}{N}} s[k \bmod M, l \bmod N]
\end{equation}
where $k, l, n, m \in \mathbb{Z}$. The \gls{dt} realization of the discrete \gls{dd} signal $s_{\mathrm{dd}}[k,l]$ is obtained by applying the Inverse Discrete Zak transform (IDZT) (see Chapter 8 in \cite{saif_book})
\begin{equation}
s[q]=\operatorname{IDZT}\left(s_{\mathrm{dd}}[k, l]\right)=\frac{1}{\sqrt{N}} \sum_{l=0}^{N-1} s_{\mathrm{dd}}[k, l],
\end{equation}
 where $q \in \mathbb{Z}$. The continuous time representation of the \gls{dt} signal $s[q]$ is given by
\begin{equation}
\widetilde{s}(t) \triangleq \sqrt{T} \sum_{q \in \mathbb{Z}} s[q] \delta(t-q / B),
\end{equation}
where $T$ and $B$ are the time and bandwidth occupied by the signal $\widetilde{s}(t)$. From \cite{saif_book}, the Zak-\gls{otfs} \gls{td} transmit signal $s(t)$ is generated by applying a pulse shaping to $\widetilde{s}(t)$ which is given by
\begin{equation}
\label{eqn:tx_signal}
s(t)=\sqrt{T} w_{1}(t) * \left[W_{2}(t) \sum_{q \in \mathbb{Z}} x[q] \delta(t-q / B)\right],
\end{equation}
where `$*$' denotes convolution. Generally in \gls{zotfs} modulation, the \gls{dd} domain input signal is passed through  \gls{dd} filter, $w_{tx}(\tau,\nu) = w_{1}(\tau)w_{2}(\nu))$, before applying the Zak transform. Therefore, in the above expression, $W_2(t)$ is the Fourier transform of the Doppler filter $w_{2}(\nu)$ and $w_1(t)$ is simply the delay filter $w_1(\tau)$ represented in time. Also, note that, $W_{2}(t)$ limits the time duration and $w_{1}(t)$ limits the bandwidth occupied by the Zak-\gls{otfs} \gls{td} transmit signal $s(t)$ (see \cite{ubadah2024,saif_book} for details).  
\subsection{Receiver Signal Processing}
\label{subsection:zak_rx}
\begin{figure}[htb!]
    \centering
    \includegraphics[width=\linewidth]{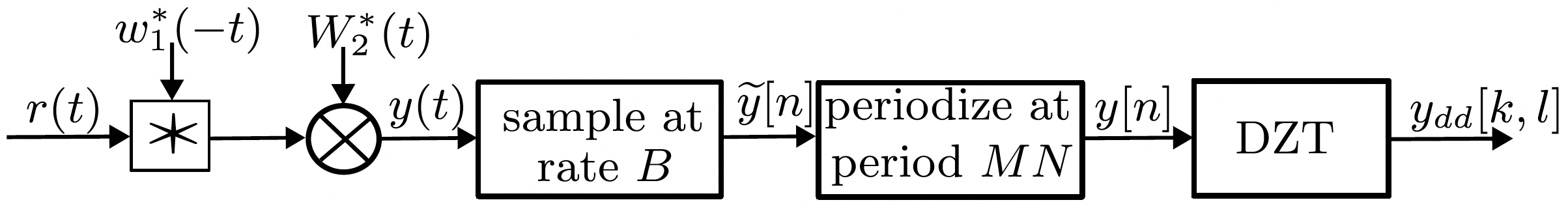}
    \caption{Zak-OTFS receiver}
    \label{fig:zak_rx}
    \vspace{-3mm}
\end{figure}
The Zak-\glsadd{otfs} receiver signal processing is shown in Fig. \ref{fig:zak_rx}. At the receiver, the received Zak-\gls{otfs} \gls{td} signal $r(t)$ is first convolved with the filter $w_1^*(-t)$ and the resultant signal is multiplied with the pulse $W^*(t)$ such that the receiver pulse shaping is matched with the transmitter (see \cite{ubadah2024, saif_book} for more details). The signal after matched filtering is given by
\begin{equation}
\label{eqn:rx_signal}
y(t)=\frac{1}{\sqrt{T}} W_{2}^{*}(t)\left[w_{1}^{*}(-t) \star r(t)\right] 
\end{equation}
The \gls{td} signal $y(t)$ is sampled at $t=\widetilde{q} / B, \widetilde{q} \in \mathbb{Z}$ to get the $
\tilde{y}[\widetilde{q}]=y(t=\widetilde{q} / B), \widetilde{q} \in \mathbb{Z}$. This signal is periodized with period $M N$ to get the DT periodic signal 
\begin{equation}
y[\widetilde{q}]=\sum_{p \in \mathbb{Z}} \tilde{y}[\widetilde{q}+p M N], \widetilde{q} \in \mathbb{Z},
\end{equation}
 which is required to generate the received \gls{dd} signal. Applying the \gls{dzt} to $y[\widetilde{q}]$ gives the received \gls{dd} signal
\begin{equation}
  y_{\mathrm{dd}}[k, l]=\operatorname{DZT}(y[\widetilde{q}])=\sum_{\tilde{n}=0}^{N-1} y[k+\tilde{n} M] \frac{e^{-j 2 \pi \tilde{n} l / N}}{\sqrt{N}}, k, l \in \mathbb{Z}   
\end{equation}
\begin{figure}[htb]
    \centering
    \vspace{-9mm}
    \includegraphics[width=0.8\linewidth]{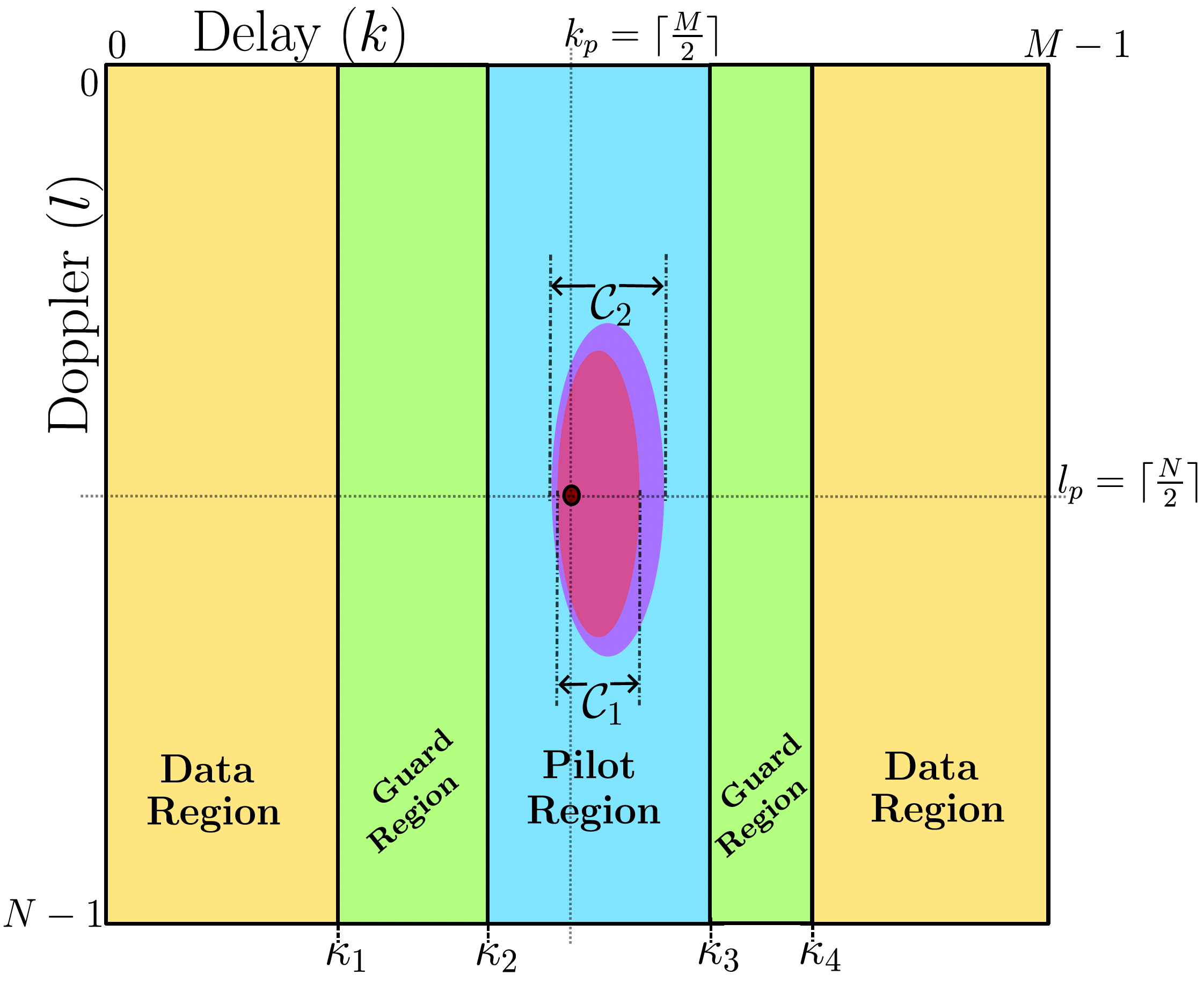}
    \caption{Zak-\gls{otfs} \gls{dd} frame with embedded pilot and data where $\kappa_1 = k_p-1-\lceil{B(\tau_{max}+\Delta t)\rceil}, \,\ \kappa_2 = k_p-1, \,\ \kappa_3 = k_p+\lceil{B(\tau_{max}+\Delta t)\rceil}, \text{ and }\kappa_4 = k_p+1+\lceil{B(\tau_{max}+\Delta t)\rceil}$.}
    \label{fig:dd_frame_struct}
\end{figure} \vspace{-2mm}

\section{\gls{ota} Implementation of Zak-\gls{otfs}}
\label{section:Otfs implementation}
For the \gls{ota} implementation of Zak-\gls{otfs} we employ the \gls{dd} frame structure shown in Fig. \ref{fig:dd_frame_struct}. The \gls{dd} frame is divided into three regions: \textit{Data region} to multiplex information symbols, \textit{Pilot region} to predict the \gls{io} relation and \textit{Guard region} to separate and avoid interference between data and pilot. To predict the \gls{io} relation, a pilot symbol is transmitted at $(k_p,l_p) = (\lceil{M/2\rceil}, \lceil{N/2\rceil})$. While the pilot region occupies all resources along the Doppler domain, its occupancy along the delay domain depends on the amount of delay spread caused by the channel \cite{Jinu_Gauss}.  From \cite{saif_book}, the received \gls{td} signal is given by
\begin{eqnarray}
\label{eqn:y_with_channel}
    r(t) &=& \iint h(\tau,\nu) s(t-\tau) e^{j2\pi \nu (t-\tau)} d\tau d\nu + z(t)
\end{eqnarray}
where $h(\tau,\nu)$ is the doubly dispersive channel, which is given by \vspace{-5mm}
\begin{eqnarray}
\label{eqn:dd_channel}
        h(\tau,\nu) &\Define& \sum_{i=1}^{p} h_i \delta(\tau-\tau_i) \delta(\nu - \nu_i)
\end{eqnarray}
where $h_i, \tau_i$ and $\nu_i$ are, respectively, the complex channel gain, delay spread and Doppler spread of the $i$-th path, and $p$ is the number of paths. Note that,  $\tau_i \le  \tau_{max}$ and $\nu_i \leq \nu_{max}$, where $\tau_{max}$ and $\nu_{max}$ are, respectively, the maximum delay and Doppler spreads of the channel. Substituting (\ref{eqn:dd_channel}) in (\ref{eqn:y_with_channel}) gives the received \gls{td} signal
\begin{eqnarray}
\label{eqn:rt_channel}
 r(t) &=& \sum_{i=1}^{p} h_i s(t-\tau_i) e^{j2\pi \nu_i (t-\tau_i)}
\end{eqnarray}

\subsection{Impact of CFO and Timing Offsets}
\label{subsec:ImpactsofCFO}
In practical wireless systems, a received signal inevitably experiences impairments such as timing offsets and carrier frequency offsets (CFO), which can significantly degrade system performance. The received \gls{td} signal $\tilde{r}(t)$ with all these practical impairments is given by
\begin{equation}
\label{eqn:rt_pi}
\tilde{r}(t) = r(t - \Delta t) \cdot e^{j(2\pi \epsilon_{0} t+ \phi(t))} + z(t)
\end{equation}
 where $\Delta t$ is the (possibly fractional) timing offset, $\epsilon_{0}$ is the normalized carrier frequency offset (CFO),  $\phi(t)$ is the phase noise process (e.g., modeled as a Wiener process), and $z(t)$ is the additive white Gaussian noise. This expression captures the cumulative effect of key physical layer impairments in practical wireless communication systems. Considering the limitations of our analysis framework, the phase noise in this work is assumed to remain constant rather than being modeled as a Wiener process\footnote{While a more accurate model represents phase noise as a Wiener process, in this work we assume constant phase noise to emphasize the effects of delay and Doppler spreading introduced by timing offsets and CFO.} i.e., $\phi(t) \approx \phi$. Substituting (\ref{eqn:rt_channel}) in (\ref{eqn:rt_pi}) gives the received signal 
 \begin{equation}
   \label{eqn:rt_pt_chanel}
\begin{aligned}
         \tilde{r}(t) =  \sum_{i=1}^{p} h_i e^{j(2\pi \epsilon_{0}(\tau_i + \Delta t) + \phi ) } s(t - \Delta t-\tau_i) \\ e^{j2\pi (\nu_i+\epsilon_{0})(t - \Delta t-\tau_i)} 
\end{aligned}  
 \end{equation}
By comparing equations (\ref{eqn:rt_channel}) and (\ref{eqn:rt_pt_chanel}), it becomes evident that the presence of \gls{cfo} and timing offsets manifests as additional Doppler and delay spreads. Specifically, the received signal can be expressed as
\begin{equation}
\label{eqn:rt_pt_chanel_new}
\tilde{r}(t) = \sum_{i=1}^{p} \widetilde{h}_i s(t - \widetilde{\tau}_i) e^{j2\pi \widetilde{\nu}_i(t - \widetilde{\tau}_i)}
\end{equation}
where the effective channel parameters are defined as follows:
$\widetilde{h}_i \triangleq h_i e^{j(2\pi \epsilon{0}(\tau_i + \Delta t) + \phi)}, \,\ \widetilde{\tau}_i = \tau_i + \Delta t$, and
$\widetilde{\nu}_i = \nu_i + \epsilon_0$. Here, $\widetilde{h}_i$, $\widetilde{\tau}_i$, and $\widetilde{\nu}_i$ represent the effective channel gain, delay, and Doppler shift, respectively, for the $i$-th propagation path in the presence of CFO and timing offset. The \gls{io} relation in (\ref{eqn:rt_pt_chanel_new}) implies that all impairments remain expressed through convolution, which can be acquired in Zak-OTFS but not in MC-OTFS. Therefore, the effect of the impairments can be inherently estimated within the effective channel itself and  eliminates the need for separate estimation procedures for CFO and timing offsets.
\subsection{Synchronization and \gls{io} prediction} 
Generally in an operational communication system, we append a preamble to a transmitted signal in order to synchronize and mitigate the effects of \gls{cfo} and timing offsets before processing the signal at the receiver. In 5G NR, particularly in \gls{ofdm} based communication systems, the  initial estimation and compensation of symbol timing and carrier frequency offset are performed using the \gls{pss} and \gls{sss} within the \gls{ssb}, with further refinement provided by the \gls{pbch} and its associated demodulation reference signals; subsequently, fine-grained frequency and timing tracking is often maintained using configured \gls{csi-rs} in connected operation \cite{3gpp38211,3gpp38212,3gpp38213}. To synchronize data and mitigate the effects of \gls{cfo} and timing offsets, the prior works on \gls{ota} implementations \cite{Zakotfs_thz, Zakotfs_sub6ghz} used a three-segmented \gls{zc} sequence based preamble with different lengths as the header. This preamble is designed to perform both coarse and fine corrections of \gls{cfo} and timing offsets. The drawback of this preamble is that it is a large sequence compared to the tranmsit signal. 
\par In this work, we propose a new synchronization scheme that employs a \gls{zc} preamble of shorter length for primary synchronization. The preamble enables accurate timing offset estimation by exploiting the excellent autocorrelation properties of the \gls{zc} sequence. In addition, a coarse estimate of the \gls{cfo} is obtained using Kay’s \gls{cfo} estimation method \cite{kay2002fast}. The received signal is then corrected for both timing and \gls{cfo} impairments before further processing. Refined estimates of these offsets are subsequently derived from the channel estimation procedure, based on the formulation in equation (\ref{eqn:rt_pt_chanel}), as discussed in the following section.
From \cite{Saif2_base}, the \gls{dd} \gls{io} relation is given by
\begin{equation}
y_{\mathrm{dd}}[k, l]=\sum_{k^{\prime}, l^{\prime} \in \mathbb{Z}} h_{\mathrm{eff}}\left[k^{\prime}, l^{\prime}\right] s_{\mathrm{dd}}\left[k-k^{\prime}, l-l^{\prime}\right] e^{j 2 \pi \frac{\left(k-k^{\prime}\right) l^{\prime}}{M N}}
\end{equation}
where $h_{\mathrm{eff}}[k,l]$ is the effective channel defined in \cite{Saif2_base}. 
The \gls{io} relation can be predicted by estimating the effective channel coefficients $h_{\mathrm{eff}}[k,l]$ from the pilot region of the \gls{dd} frame (see Fig. \ref{fig:dd_frame_struct}). As described in \cite{Saif2_chan}, the estimate of the effective channel $\widehat{h}_{\mathrm{eff}}[k,l]$  is given by 
 \begin{equation}
 \label{eqn:channelestimate}
     \widehat{h}_{\text {eff }}[k, l]=\left\{\begin{array}{cc} \hspace{-2mm}y_{\mathrm{dd}}\left[k+\frac{M}{2}, l+\frac{N}{2}\right] e^{-j \pi \frac{l}{N}},   &  \kappa_2 \leq k<\kappa_3, \\ & -\frac{N}{2} \leq l<\frac{N}{2} \\ \hspace{-30mm} 0, & \text {otherwise.}\end{array}\right.
 \end{equation}
From Fig. \ref{fig:dd_frame_struct} and (\ref{eqn:channelestimate}), it is clear that if the \gls{cfo} and timing offsets are fully corrected, then the support of the effective channel is $\mathcal{C}_1$. If they are not fully corrected then the support of the effective channel is $\mathcal{C}_2$. This means that the refined estimates of the offsets are inherently estimated within the effective channel by choosing the support $\mathcal{C}_2$. Using this predicted \gls{io}, we utliized \gls{mmse} equalization to mitigate the effect of the channel and estimate the information symbols. 
\begin{figure}[h]
    \centering
    \includegraphics[width=0.9\linewidth]{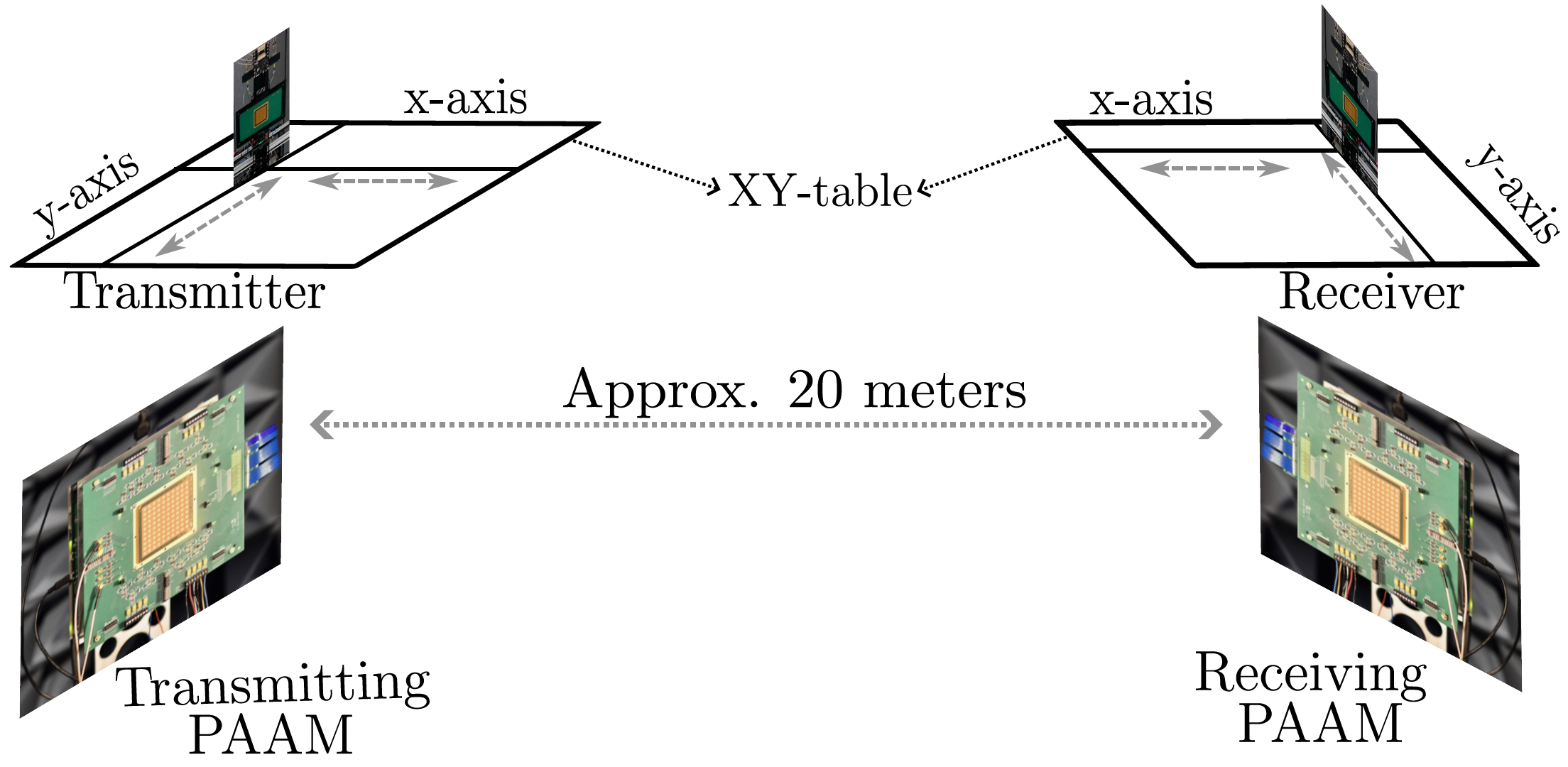}
    \caption{\gls{mmwave} experimental setup featuring a 64 PAAM transmitter and receiver mounted on an XY-table with controlled mobility, enabling motion equivalent to pedestrian walking speeds.}
    \label{fig:mmwave_setup}
    \vspace{-6mm}
\end{figure} 
\subsection{\gls{ota} mmWave Experimental Setup}
Our \gls{ota} mmWave experimental setup is shown in Fig. \ref{fig:mmwave_setup}. For the experiments, we are using the COSMOS testbed \cite{cosmos2025} with
IBM $28$ GHz $64$-element dual-polarized phased array antenna module (PAAM) which is integrated with a USRP N310  \cite{chen2022programmable}. As shown in Fig. \ref{fig:mmwave_setup}, we mount the \gls{mmwave} transceiver on a XY-table that allows each \gls{mmwave} front end to move independently in the horizontal plane within an area of $1.3 \times 1.3$ meters. 

The \gls{zotfs} transceiver processing is implemented in C++/Python and interfaced with a USRP N310 software-defined radio. A signal generated by the custom C++/Python \gls{zotfs} implementation is upsampled and upconverted from the \gls{if} to the mmWave band before transmission on the COSMOS testbed. At the receiver, located approximately 20 meters from the transmitter, the mmWave signal is downconverted to \gls{if} and returned to a USRP N310, where it is downsampled and processed using the custom Zak-OTFS receiver for performance analysis.






 \section{Results}
 \label{section:results}

The experimental results for the \gls{zotfs} system discussed in Section \ref{section:Otfs implementation} are performed at a \gls{mmwave} frequency of 28 GHz. For initial synchronization, we used a \gls{zc} based preamble of length 256 samples and observed that this short preamble is enough to detect the start of the signal and mitigate the effect of timing offsets. For the experiments performed in this work, we observe that the short preamble suffices to estimate the \gls{cfo} correctly. But it does not impact the performance of our design because of the robustness of \gls{zotfs} to \gls{cfo} (more details are mentioned in the subsequent sections). The $\kappa_i, (i=1,2,3,4)$ parameters of the \gls{dd} frame are chosen randomly based on the XY table movement and whether \gls{cfo} correction is performed at \gls{td} or not. Also, the \gls{dd} domain filter used is \gls{rrc} where its corresponding \gls{td} representation (see (\ref{eqn:tx_signal}) and (\ref{eqn:rx_signal})) is given at the top on next page in (\ref{eqn:filter}). Also, in this paper, for the \gls{rrc} filter we used $\beta=0.5$.
Other \gls{zotfs} and  hardware parameters are listed in Table \ref{table:parameters}. For the experimental results described further, we varied the USRP Tx/Rx gains such that the recieved SNR is approximately 25 dB. The following subsections present the observations and experimental validations conducted in this work. 
\begin{table}[htbp]
\vspace{-3mm}
\caption{Zak-\gls{otfs} \& Hardware Parameters}
\begin{center}
\vspace{-3mm}
\begin{tabular}{|c|c|}
\hline
Parameters & Values \\
\hline
Base Modulation & 4/16 QAM \\
Doppler Period ($\nu_p$) & 30~kHz\\
Delay Period ($\tau_p = 1/\nu_p$) & 33.3 $\mu$s \\
Doppler Taps ($N$) & 64\\
Delay Taps ($M$) & 64\\
Bandwidth ($B$) & 1.92~MHz\\
USRP Tx/Rx Gain & Varied\\
Tx/Rx Separation & 20~m\\
Sampling rate & 15.625~Msps \\
IF Frequency & 3~GHz\\
RF Frequency & 28~GHz\\
\hline
\end{tabular}
\label{table:parameters}
\end{center}
\vspace{-5mm}
\end{table} 
\begin{figure}[htb!]
\vspace{-2mm}
    \centering
 \includegraphics[width=0.35\textwidth]{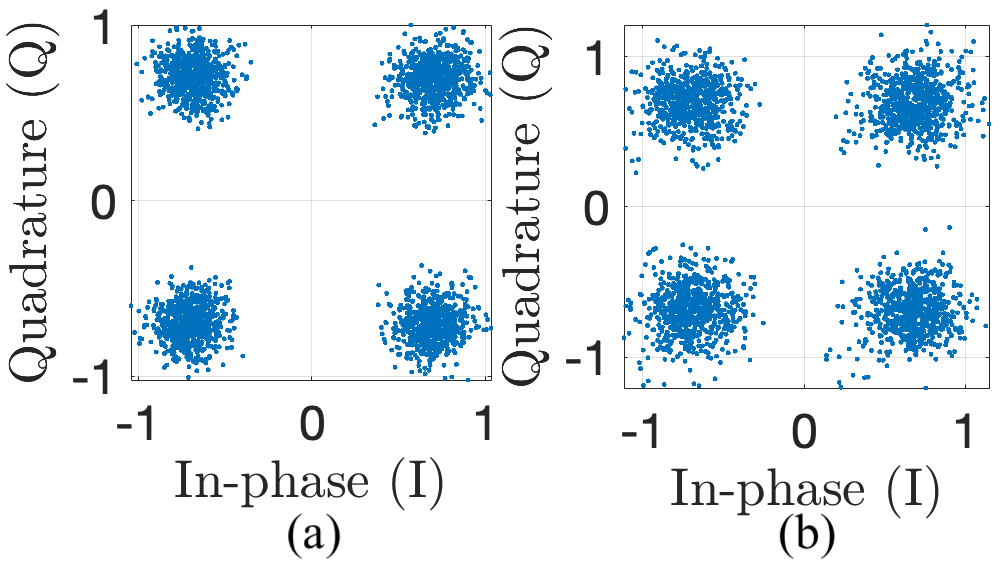}
 \caption{4-QAM Received constellation points for different scenarios. (a) RRC filter with CFO correction and stationary XY-table, (b) RRC filter without CFO correction and with movement of XY-table.}
 \label{fig:result_1}
 \vspace{-8mm}
\end{figure}
\begin{figure}[htb!]
    \centering
    \includegraphics[width=\linewidth]{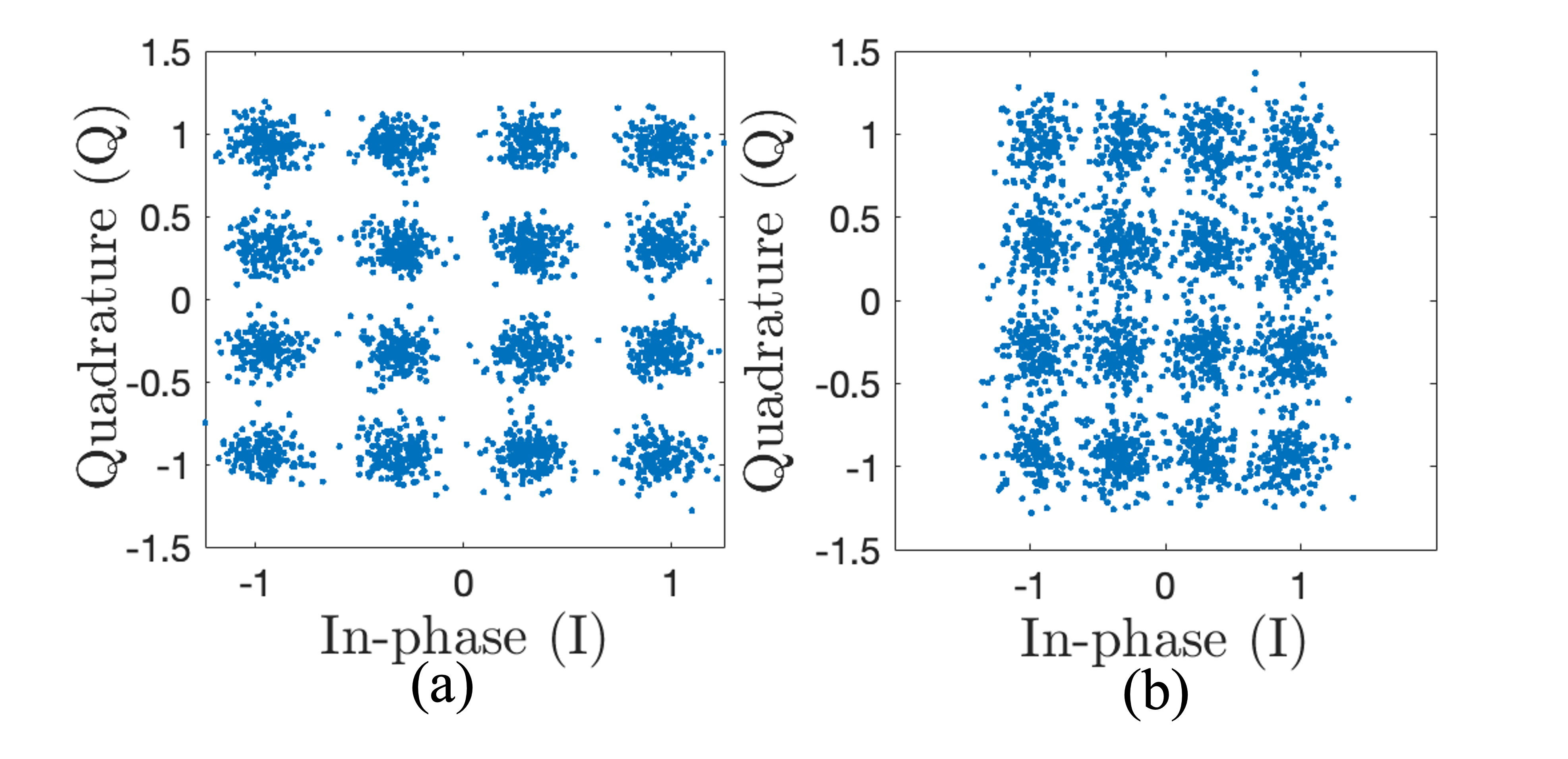}
    \caption{16-QAM Received constellation points with transceivers moving on the XY table. (a) RRC filter with CFO correction, (b) RRC filter without CFO correction.}
    \label{fig:16qam}
    \vspace{-6mm}
\end{figure}
\begin{figure*}[t]
\begin{eqnarray}
\label{eqn:filter}
   \hspace{-5mm}& & \hspace{-10mm} w_1(t) =
\frac{ \sin\left( \pi Bt (1 - \beta) \right) + 4 \beta Bt \cos\left( \pi Bt (1 + \beta) \right) }
{ \pi Bt \left[ 1 - (4 \beta Bt)^2 \right] }, 
W_2(t) = 
\begin{cases}
\frac{1}{\sqrt{T}}, & |t| \leq t_1 \\
\sqrt{ \dfrac{1}{2T} \left[ 1 + \cos\left( \dfrac{\pi}{\beta T} \left( |t| - \dfrac{(1 - \beta)T}{2} \right) \right) \right] }, & t_1 < |t| \leq t_2 \\
0, & |t| > t_2, 
\end{cases} \nonumber \\
& & \qquad \qquad \qquad \qquad \qquad \qquad \qquad \qquad \qquad \qquad \qquad \qquad \text{where } t_1 = \frac{(1 - \beta)T}{2} \text{ and } t_2 = \frac{(1 + \beta)T}{2}.  
\end{eqnarray}
\vspace{-3mm}
\end{figure*}
\begin{figure}[htb!]
\vspace{-5mm}
    \centering
    \includegraphics[width=0.75\linewidth]{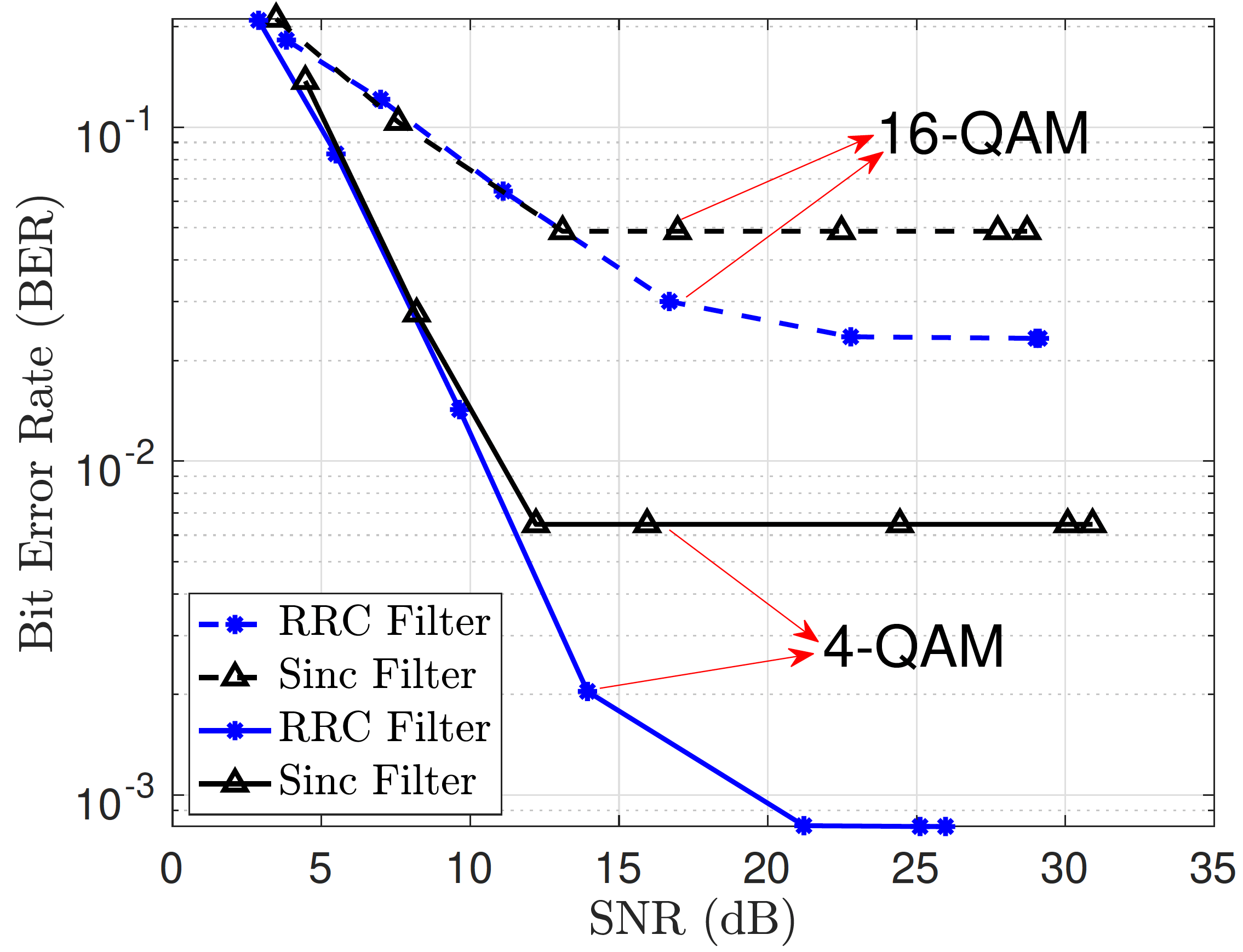}
    \caption{BER versus SNR curve with CFO correction and transceivers moving on the XY table}
    \label{fig:ber}
    \vspace{-7mm}
\end{figure}
\subsection{Robustness of Zak-OTFS to CFO}
Fig. \ref{fig:result_1} shows the received 4-QAM constellations for different scenarios. Fig. \ref{fig:result_1} (a) shows the received constellation when the XY-table is stationary, where the system employs a \gls{rrc} filter, and \gls{cfo} corrections is applied in the \gls{td} before processing the signal in the \gls{dd} domain.  Similarly, Fig. \ref{fig:result_1} (b) shows the received constellation when the XY-table is moving, where the system employs a \gls{rrc} filter, and \gls{cfo} corrections are not applied in the \gls{td}. Fig. \ref{fig:result_1} clearly shows the impact of movement and \gls{cfo} on the received constellation. The impact can be seen as the spreading of the constellation rather than rotation or shifting.  The other reason is related to the imperfect choice of the support set $\mathcal{C}_1$ in the estimation of the \gls{io} relation. For \ref{fig:result_1}, to make the observations clear and consistency with the analysis in Section \ref{subsec:ImpactsofCFO},  we predicted the \gls{io} relation only in support $\mathcal{C}_1$ for both the cases. The performance in \ref{fig:result_1} (b) can be further improved by selecting the support $\mathcal{C}_2$ for predicting the \gls{io} relation. Overall, we observed that the I/O relation given by convolution (see (\ref{eqn:rt_pt_chanel_new})) simplifies the \gls{ota} implementation, because \gls{cfo} and timing offsets can, to some degree, be folded into the estimation of the effective channel.
\subsection{Performance with Moving Transceivers}
Fig. \ref{fig:16qam} shows the received 16-QAM constellations when the transceivers are moving over the XY table. Fig. \ref{fig:16qam} (a) shows the received constellation for the case where initial \gls{cfo} estimation and correction is performed in the \gls{td} followed by estimation and equalization of the effective channel. However, for the received constellation in Fig. \ref{fig:16qam} (b), rather than estimating and correcting the \gls{cfo} in the \gls{td}, we used our analysis in Section \ref{section:Otfs implementation} and inherently estimated the effects of \gls{cfo} within the effective channel by choosing an appropriate support region $\mathcal{C}_2$. It is observed from Fig. \ref{fig:16qam} that the constellation spreads more for the case where the correction \gls{cfo} is not performed in \gls{td}, but selecting a good support region $\mathcal{C}_2$ can increase the precision to predict the relation \gls{io} and thus reduce the spread of the constellation.  The choice of an appropriate support region $\mathcal{C}_2$ will obtain the spread caused by \gls{cfo} within the estimate of the effective channel. This is not the case with OFDM where the \gls{cfo} results in \gls{ici} which disturbs the orthogonality of the waveform. 

Fig. \ref{fig:ber} shows the BER versus SNR curve for different filters with CFO correction in \gls{td} and transceivers moving on the XY table. It can be observed from Fig. \ref{fig:ber} that the uncoded BER performance of Zak-OTFS for both 4 and 16-QAM  is better when a \gls{rrc} filter is used compared to that of a sinc filter. Also, the BER performance observed in Fig. \ref{fig:ber} is consistent with the developed theory in \cite{Jinu_Gauss}. 

\section{Conclusion}
\label{section:conclusion}
This paper presented one of the first \gls{ota} implementations of \gls{zotfs} modulation at \gls{mmwave} frequencies. The proposed system demonstrated improved input–output predictability using \gls{rrc} filtering, efficient synchronization through a low-overhead preamble, and robust operation under \gls{cfo} and timing impairments. Experimental results validated the feasibility and robustness of \gls{zotfs} in realistic \gls{mmwave} conditions, confirming its potential as a promising physical-layer waveform for beyond-5G and 6G systems. In future work, the system will be enhanced for real-time operation with reduced latency and computational complexity. Furthermore, outdoor experiments with mobile transmitter and receiver setups are planned to evaluate \gls{zotfs} performance under high-speed motion, enabling observation of significant Doppler effects and further validation in dynamic propagation environments. \vspace{-2mm}
\section*{Acknowledgment}
The Duke team is supported by the US National Science Foundation (NSF) (grants 2342690,  2342690, and 214821), in-part by the Air Force Office of Scientific Research (grants FA8750-20-2-0504 and FA9550-23-1-0249), and in-part by federal agency and industry partner funds as specified in the Resilient \& Intelligent NextG Systems (RINGS) program.
Approved for Public Release; Distribution Unlimited: AFRL-2025-1371. The Rutgers team is supported by the NSF under grants 2128077 and 2450567.
\bibliographystyle{ieeetr}
\bibliography{References}

\end{document}

%% file: acronyms.tex
\newacronym{sdr}{SDR}{software-defined radio}
\newacronym{iq}{IQ}{in-phase and quadrature}
\newacronym{if}{IF}{intermediate frequency}
\newacronym{lna}{LNA}{low-noise amplifier}
\newacronym{rf}{RF}{radio frequency}
\newacronym{6g}{6G}{sixth generation}
\newacronym{vr}{VR}{virtual reality}
\newacronym{adc}{ADC}{analog to digital converter}
\newacronym{dac}{DAC}{digital to analog converter}
\newacronym{awg}{AWG}{arbitrary waveform generator}
\newacronym{dso}{DSO}{digital storage oscilloscope}
\newacronym{psg}{PSG}{performance signal generator}
\newacronym{thz}{THz}{terahertz}
\newacronym{sub-thz}{sub-THz}{sub-terahertz}
\newacronym{fft}{FFT}{fast Fourier transform}
\newacronym{css}{CSS}{chirp spread spectrum}
\newacronym{ofdm}{OFDM}{Orthogonal Frequency Division Multiplexing}
\newacronym{lo}{LO}{local oscillator}
\newacronym{otfs}{OTFS}{Orthogonal Time Frequency Space}
\newacronym{papr}{PAPR}{peak-to-average power ratio}
\newacronym{mmse}{MMSE}{minimum mean squared error}
\newacronym{ber}{BER}{bit error rate}
\newacronym{vdi}{VDI}{Virginia Diodes Inc.}
\newacronym{awgn}{AWGN}{additive white Gaussian noise}
\newacronym{dft}{DFT}{discrete Fourier transform}
\newacronym{psk}{PSK}{phase-shift keying}
\newacronym{qam}{QAM}{quadrature amplitude modulation}
\newacronym{evm}{EVM}{error vector magnitude}
\newacronym{bpsk}{BPSK}{binary phase-shift keying}
\newacronym{qpsk}{QPSK}{quadrature phase-shift keying}
\newacronym{ifft}{IFFT}{inverse fast Fourier transform}
\newacronym{sfft}{SFFT}{symplectic finite Fourier transform}
\newacronym{isfft}{ISFFT}{inverse symplectic finite Fourier transform}
\newacronym{idft}{IDFT}{inverse discrete Fourier transform}
\newacronym{cfo}{CFO}{carrier frequency offset}
\newacronym{cpo}{CPO}{carrier phase offset}
\newacronym{mixamc}{MixAMC}{mixer/amplifier/multiplier-chain}
\newacronym{usrp}{USRP}{universal software radio peripheral}
\newacronym{dd}{DD}{delay-Doppler}
\newacronym{tx}{Tx}{transmitter}
\newacronym{rx}{Rx}{receiver}
\newacronym{los}{LOS}{line-of-sight}
\newacronym{snr}{SNR}{signal-to-noise ratio}
\newacronym{td}{TD}{time domain}
\newacronym{2d}{2D}{two dimensional}
\newacronym{lpf}{LPF}{low-pass filter}
\newacronym{dzt}{DZT}{discrete time Zak transform}
\newacronym{ml}{ML}{maximum likelihood}
\newacronym{isac}{ISAC}{integrated sensing and communications}
\newacronym{aiml}{AI/ML}{artificial intelligence/machine learning}
\newacronym{io}{I/O}{input-output}
\newacronym{ota}{OTA}{Over-The-Air}
\newacronym{dt}{DT}{discrete-time}
\newacronym{pbch}{PBCH}{Physical Broadcast Channe}
\newacronym{csi-rs}{CSI-RS}{channel state information reference signal}
\newacronym{ssb}{SSB}{Synchronization Signal Block}
\newacronym{sss}{SSS}{Secondary Synchronization Signa}
\newacronym{pss}{PSS}{Primary Synchronization Signal}
\newacronym{zc}{ZC}{Zadoff-Chu}
\newacronym{mmwave}{mmWave}{Millimeter wave}
\newacronym{tf}{TF}{time frequency}
\newacronym{ici}{ICI}{intercarrier interference}
\newacronym{zotfs}{Zak-OTFS}{Zak transform}
\newacronym{mcotfs}{MC-OTFS}{multicarrier OTFS}
\newacronym{rrc}{RRC}{root-raised-cosine}

%% file: References.bib
@INPROCEEDINGS{Hadani2017,
  author={Hadani, R. and Rakib, S. and Tsatsanis, M. and Monk, A. and Goldsmith, A. J. and Molisch, A. F. and Calderbank, R.},
  booktitle={2017 IEEE Wireless Communications and Networking Conference (WCNC)}, 
  title={Orthogonal Time Frequency Space Modulation}, 
  year={2017},
  volume={},
  number={},
  pages={1-6},
  doi={10.1109/WCNC.2017.7925924}}

@ARTICLE{Saif2_base,
  author={Mohammed, Saif Khan and Hadani, Ronny and Chockalingam, Ananthanarayanan and Calderbank, Robert},
  journal={IEEE BITS the Information Theory Magazine}, 
  title={{OTFS}—A Mathematical Foundation for Communication and Radar Sensing in the Delay-Doppler Domain}, 
  year={2022},
  volume={2},
  number={2},
  pages={36-55},
  doi={10.1109/MBITS.2022.3216536}}

@ARTICLE{Saif2_chan,
  author={Mohammed, Saif Khan and Hadani, Ronny and Chockalingam, Ananthanarayanan and Calderbank, Robert},
  journal={IEEE BITS the Information Theory Magazine}, 
  title={{OTFS}—Predictability in the Delay-Doppler Domain and Its Value to Communication and Radar Sensing}, 
  year={2023},
  volume={3},
  number={2},
  pages={7-31},
  doi={10.1109/MBITS.2023.3319595}}

@misc{ubadah2024,
      title={Zak-OTFS to Integrate Sensing the I/O Relation and Data Communication}, 
      author={Muhammad Ubadah and Saif Khan Mohammed and Ronny Hadani and Shachar Kons and Ananthanarayanan Chockalingam and Robert Calderbank},
      year={2025},
      eprint={2404.04182},
      archivePrefix={arXiv},
      primaryClass={eess.SP},
      url={https://arxiv.org/abs/2404.04182}, 
}

@book{saif_book,
  title     = "OTFS Modulation: Theory and Applications",
  author    = "Saif Khan Mohammed and Ronny Hadani and Ananthanarayanan Chockalingam",
  year      = 2024,
  publisher = "Wiley-IEEE Press",
  isbn   = "9781119984214"

}

@misc{Jinu_Gauss,
      title={Zak-OTFS: Pulse Shaping and the Tradeoff between Time/Bandwidth Expansion and Predictability}, 
      author={Jinu Jayachandran and Rahul Kumar Jaiswal and Saif Khan Mohammed and Ronny Hadani and Ananthanarayanan Chockalingam and Robert Calderbank},
      year={2024},
      eprint={2405.02718},
      archivePrefix={arXiv},
      primaryClass={eess.SP},
      url={https://arxiv.org/abs/2405.02718}, 
}

@techreport{3gpp38211,
  title        = {{3rd Generation Partnership Project; Technical Specification Group Radio Access Network; NR; Physical channels and modulation (Release 17)}},
  institution  = {3rd Generation Partnership Project (3GPP)},
  number       = {TS 38.211},
  year         = {2024},
  month        = {March},
  note         = {\url{https://www.3gpp.org/DynaReport/38211.htm}},
}

@techreport{3gpp38212,
  title        = {{3rd Generation Partnership Project; Technical Specification Group Radio Access Network; NR; Multiplexing and channel coding (Release 17)}},
  institution  = {3rd Generation Partnership Project (3GPP)},
  number       = {TS 38.212},
  year         = {2024},
  month        = {March},
  note         = {\url{https://www.3gpp.org/DynaReport/38212.htm}},
}

@techreport{3gpp38213,
  title        = {{3rd Generation Partnership Project; Technical Specification Group Radio Access Network; NR; Physical layer procedures for control (Release 17)}},
  institution  = {3rd Generation Partnership Project (3GPP)},
  number       = {TS 38.213},
  year         = {2024},
  month        = {March},
  note         = {\url{https://www.3gpp.org/DynaReport/38213.htm}},
}

@misc{Zakotfs_sub6ghz,
    title={{Zak-OTFS} with Spread Pilot in Sub-6 {GHz}: Implementation and Over-the-Air Experimentation},
  author={Junyao Zheng and Venkatesh Khammammetti and Beyza Dabak and Sandesh Rao Mattu and Tingjun Chen and Robert Calderbank},
  note={Accepted for publication in the IEEE Military Communications Conference (MILCOM)},
  year={2025},
}

@misc{Zakotfs_thz,
      title={Over-the-Air Transmission of Zak-OTFS with Spread Pilots on Sub-THz Communications Testbed}, 
      author={Claire Parisi and Venkatesh Khammammetti and Robert Calderbank and Lauren Huie},
      year={2025},
      eprint={2504.15947},
      archivePrefix={arXiv},
      primaryClass={eess.SP},
      note = {\url{https://arxiv.org/abs/2504.15947}}, 
}

@article{mmwave5G_2,
  title     = {Millimetre wave frequency band as a candidate spectrum for 5G network},
  author    = {S. Rangan and T. S. Rappaport and E. Erkip},
  journal   = {Procedia Computer Science},
  volume    = {109},
  pages     = {776--781},
  year      = {2017},
  publisher = {Elsevier},
  url       = {https://www.sciencedirect.com/science/article/pii/S1874490717305827},
  note      = {Accessed: 2025-10-07}
}

@article{mmwave_challenges_1,
  title     = {Millimeter Wave Mobile Communications for 5G Cellular: It Will Work!},
  author    = {Theodore S. Rappaport and Shu Sun and Rimma Mayzus and Hang Zhao and Yaniv Azar and Kostas Wang and George N. Wong and Jocelyn K. Schulz and Mathew Samimi and Felix Gutierrez},
  journal   = {IEEE Access},
  volume    = {1},
  pages     = {335--349},
  year      = {2013},
  publisher = {IEEE},
  doi       = {10.1109/ACCESS.2013.2260813},
  url       = {https://doi.org/10.1109/ACCESS.2013.2260813},
  note      = {Discusses path loss, blockage, and mobility challenges in mmWave systems.}
}

@article{mmwave_challenges_2,
  title     = {Millimeter-Wave Cellular Wireless Networks: Potentials and Challenges},
  author    = {Sundeep Rangan and Theodore S. Rappaport and Elza Erkip},
  journal   = {Proceedings of the IEEE},
  volume    = {102},
  number    = {3},
  pages     = {366--385},
  year      = {2014},
  publisher = {IEEE},
  doi       = {10.1109/JPROC.2014.2299397},
  url       = {https://doi.org/10.1109/JPROC.2014.2299397},
  note      = {Analyzes mmWave propagation, blockage, hardware impairments, and mobility effects.}
}

@article{OFDM_ICI_3,
  title     = {On the Performance of OFDM Systems for High-Mobility Channels},
  author    = {Changyong Wang and Yi Zhang and Shidong Zhou and Jing Wang},
  journal   = {IEEE Journal on Selected Areas in Communications},
  volume    = {27},
  number    = {6},
  pages     = {899--908},
  year      = {2009},
  publisher = {IEEE},
  doi       = {10.1109/JSAC.2009.090718},
  url       = {https://doi.org/10.1109/JSAC.2009.090718},
  note      = {Evaluates OFDM under time-varying channels; shows degradation due to Doppler and ICI.}
}

@techreport{Hadani2015OTFSWhitePaper,
        title={Orthogonal Time Frequency Space Modulation}, 
      author={Ronny Hadani and Shlomo Rakib and Shachar Kons and Michael Tsatsanis and Anton Monk and Christian Ibars and Jim Delfeld and Yoav Hebron and Andrea J. Goldsmith and Andreas F. Molisch and Robert Calderbank},
      year={2018},
      eprint={1808.00519},
      archivePrefix={arXiv},
      primaryClass={cs.IT},
      url={https://arxiv.org/abs/1808.00519}, 
}

@inproceedings{Samsung_mmwave,
  title     = {Performance Analysis of OTFS Waveform for 5G NR mmWave Communication System},
  author    = {Gunturu, Anusha and Godala, Anirudh Reddy and Sahoo, Ashok Kumar and Chavva, Ashok Kumar Reddy and others},
  booktitle = {IEEE Wireless Communications and Networking Conference (WCNC)},
  year      = {2021},
  pages     = {1--6},
  doi       = {10.1109/WCNC49053.2021.9417346},
  url       = {https://doi.org/10.1109/WCNC49053.2021.9417346},
  note      = {Evaluates OTFS and OFDM over 5G NR mmWave chain using 3GPP TDL/CDL models, reports BER and complexity tradeoffs.}
}

@article{OTFS_mmwave_2025,
  title     = {A Comparison among Single Carrier, {OFDM}, and {OTFS} in mmWave Multi-Connectivity Downlink Transmissions},
  author    = {Fabian Gottsch and Shuangyang Li and Lorenzo Miretti and Giuseppe Caire and Sławomir Stańczak},
  journal   = {arXiv preprint arXiv:2503.23745},
  year      = {2025},
  url       = {https://arxiv.org/abs/2503.23745},
  note      = {Accessed: 2025-10-07; comparative study in mmWave multi-connectivity scenarios among SC, OFDM, OTFS}
}

@article{OTFS_mmwave_2023_1,
  title     = {Low-complexity OTFS-based hybrid precoding in mmWave high mobility A2G systems},
  author    = {Zhang, Yujie and Wang, Xinyu and Li, Hongwei and Wang, Feng and Xu, Cheng},
  journal   = {Digital Signal Processing},
  volume    = {139},
  pages     = {104044},
  year      = {2023},
  publisher = {Elsevier},
  doi       = {10.1016/j.dsp.2023.104044},
  url       = {https://www.sciencedirect.com/science/article/pii/S1051200423001382},
  note      = {Proposes a low-complexity hybrid digital--analog precoding design for OTFS-based mmWave A2G systems.}
}

@article{OTFS_mmwave_2023_2,
  title     = {Orthogonal Time Frequency Space with Delay-Doppler Alignment Modulation (DDAM-OTFS)},
  author    = {Li, Shuangyang and Caire, Giuseppe},
  journal   = {arXiv preprint arXiv:2303.06156},
  year      = {2023},
  url       = {https://arxiv.org/abs/2303.06156},
  note      = {Introduces DDAM-OTFS with per-path delay and Doppler compensation for mmWave/THz channels.}
}

@inproceedings{Surabhi2019OTFSPhaseNoise,
  title     = {OTFS Modulation with Phase Noise in mmWave Communications},
  author    = {G. D. Surabhi and M. Kollengode Ramachandran and A. Chockalingam},
  booktitle = {Proc. IEEE Vehicular Technology Conference (VTC Spring)}, 
  year      = {2019},
  pages     = {1--5},
  doi       = {10.1109/VTCSpring.2019.8746382},
  url       = {https://ece.iisc.ac.in/~wrl/paper/Phase_Noise_OTFS.pdf},
  note      = {Incorporates phase noise into DD domain model and compares BER of OTFS vs. OFDM at 28 GHz}
}

@online{cosmos2025,
  author       = "{COSMOS Testbed}",
  title        = "{COSMOS – Cloud Enhanced Open Software Defined Mobile Wireless Testbed}",
  year         = "2025",
  url          = "https://www.cosmos-lab.org/",
  note         = "Accessed: Oct. 7, 2025"
}

@inproceedings{chen2022programmable,
  title={Programmable and open-access millimeter-wave radios in the PAWR COSMOS testbed},
  author={Chen, Tingjun and Maddala, Prasanthi and Skrimponis, Panagiotis and Kolodziejski, Jakub and Gu, Xiaoxiong and Paidimarri, Arun and Rangan, Sundeep and Zussman, Gil and Seskar, Ivan},
  booktitle={Proceedings of the 15th ACM Workshop on Wireless Network Testbeds, Experimental evaluation \& CHaracterization},
  pages={1--8},
  year={2022}
}

@article{kay2002fast,
  title={A fast and accurate single frequency estimator},
  author={Kay, Steven},
  journal={IEEE Transactions on Acoustics, Speech, and Signal Processing},
  volume={37},
  number={12},
  pages={1987--1990},
  year={2002},
  publisher={IEEE}
}
